\documentclass[preprint,pre,showpacs,epsfig,address]{revtex4}
\usepackage[english]{babel}
\usepackage[latin1]{inputenc}
\usepackage[dvips]{graphicx}
\usepackage{amsmath}
\usepackage{amssymb}
\usepackage{epsfig}
\usepackage{array}
\usepackage{amsfonts}
\usepackage{graphicx}%
\begin{document}
\title{ Base sequence dependent sliding of proteins on DNA }
\author{M. Barbi }
\affiliation{Laboratoire de Physique Th\'{e}orique des Liquides, Universit{\'{e}} Pierre et
Marie Curie, case courrier 121, 4 Place Jussieu - 75252 Paris cedex 05, France }
\author{C. Place }
\affiliation{Laboratoire de Physique, CNRS-UMR 5672, {\'{E}}cole Normale Sup{\'{e}}rieure
de Lyon, Lyon, France }
\author{V. Popkov}
\affiliation{Institut f\"{u}r Festk\"{o}rperforschung, Forschungszentrum J\"{u}lich - 52425
J\"{u}lich, Germany}
\author{M. Salerno }
\affiliation{Dipartimento di Fisica ``E.R. Caianiello" and Istituto Nazionale di Fisica
della Materia (INFM), Universit\'{a} di Salerno, I-84081 Baronissi (SA), Italy}
\date{\today }

\begin{abstract}
The possibility that the sliding motion of proteins on DNA is  influenced by
the base sequence through a base pair reading  interaction, is considered.
Referring to the case of the T7  RNA-polymerase, we show that the protein
should follow a  noise-influenced sequence-dependent motion which deviate from
the  standard random walk usually assumed. The general validity and the
implications of the results are discussed.

\end{abstract}
\maketitle

\section{Introduction}

How site-specific DNA binding proteins locate their targets on DNA is an issue
of primary importance for understanding the functioning of DNA. With the
development of new experimental techniques, this problem is getting much of
attention, see, e.g., \cite{Gut99,Shi99,Ger02,Bru02,Sta00,Hal02,GowersHalford}%
. Sliding, hopping and uncorrelated three-dimensional diffusion are generally
taken into account as possible searching mechanisms, and their relative role
in target location is being discussed and experimentally investigated. In the
seminal work of Berg, Winter and von Hippel (BWH), one-dimensional diffusion
(sliding) along DNA was proposed as a necessary ingredient of the target
search \cite{Ber81}. More recent papers \cite{Bru02,Sta00,Hal02} confirm the
importance of sliding in the search process, along with three dimensional
paths (disattachment of a protein from DNA and reattachment to a different
segment of DNA) \cite{GowersHalford}.

A completely coherent description of the search process is nevertheless still
lacking. In a recent paper \cite{Bru02}, Bruinsma remarks e.g. that the time
spent by lac-repressor on each DNA site in the frame of the BWH theory is too
short to allow the structural changes necessary for the protein to recognize
its target. He thus indicates the need for a slowing down effect and suggests
that \textquotedblleft indirect read-out\textquotedblright\ mechanisms,
associated to the DNA flexibility, can account for it. Note that the DNA
sequence, responsible for the DNA flexibility and shape, is crucial also for
this kind of slowing down effect.

On the other hand, all existing models of target search dynamics describe the
sliding motion as a \emph{standard random walk}. In  theoretical analysis of
experiments it is indeed taken for granted that the protein motion is governed
by a linear diffusion, $\langle x^{2}\rangle=2D\,t$. While the linear
diffusion assumption is natural for 3-dimensional paths (when protein is not
bound to DNA and diffuses in solution), for sliding phase of motion it implies
that the DNA is essentially \textquotedblleft seen\textquotedblright\ by the
protein as a homogeneous chain \cite{Kafri_condmat}. This homogeneity of DNA, however, seems
incompatible with the recognition function, which always involves a form of
\emph{reading}, so that it is natural to assume an influence of the DNA
sequence on the sliding dynamics. This influence could result in slowing down,
pauses and stops which, in its turn, could invalidate the random walk
assumption. These slowing effects can have have a different origin from that
suggested by Bruinsma \cite{Bru02}; note, nevertheless, that different
mechanisms can coexist, and that in any case the dynamic effects of (direct
or indirect) sequence sensitivity are considered.

The aim of the present paper is to show that sequence dependence of the
DNA-protein interaction can induce strong deviations from standard diffusion
for a generic protein sliding on DNA. To this regard, we use a probabilistic
model for the sliding motion of a protein on DNA in which the influence of the
base sequence is accounted through the DNA-protein reading interaction
\cite{Bar04}. As a result we show that the protein follows a
noise-influenced sequence-dependent motion which deviates from standard
diffusion, reaching normal diffusion only at asymptotically large times. The
presence of an anomalous diffusion (AD) regime speeds up the mobility of a
protein thus greatly facilitating the target search. The cross-over from
anomalous to normal diffusion occurs at times typically needed for a protein
to cover the distance at which the potential averages out (of order 100 bp in
our model). On the other hand, indirect measurements hint on the typical mean
path length traversed by the protein during a single DNA binding event, of the
same order of magnitude (e.g., around 170 bp in \cite{GowersHalford}). Thus,
the anomalous diffusion (AD) should actually dominate the binding phase, and
cannot be neglected.

The paper is organized as follows. In section II we introduce the model using
T7 RNA-polymerase as a specific example of a sliding protein. In
section III we investigate the main properties of the sliding dynamics
including the sub-diffusive regime and the crossover to normal diffusion. In
section IV we provide some arguments supporting the generality of our results
in connection to applications to other enzymes. Finally, in section V, 
results and  conclusions of the paper are summarized.

\section{The model}

A target sequence usually consists of few (say, $r$) consecutive base pairs
(bps). Specific sequence recognition is often mediated by hydrogen bonds
(H-bonds) to a set of four specific, spatially ordered chemical groups on the
major groove side of the bps \cite{See76,Nad99}. Besides this mechanism, other
features of DNA such as shape and flexibility, as well as electrostatic
interactions between protein and DNA \cite{Kam00,travers} may also be involved
in the recognition process. In this paper, we will focus mainly on the first
mechanism, i.e., we assume that proteins check the sequence at each position on
DNA by exploiting the same set of hydrogen bonds they form with the DNA at the
target site. We thus represent the DNA binding sites at position $n$ as a
sequence of $r$ vectors $b_{n}$ (one for each bp), of the form $B_{n}%
=(b_{n},b_{n+1},\dots,b_{n+r-1})$, according to the rule
\begin{equation}
b_{n}=\left\{
\begin{array}
[c]{lr}%
(1,-1,1,0)^{T}\,\,\,\mbox{for AT},\,\,\,\,\,\, & (0,1,-1,1)^{T}%
\,\,\,\mbox{for TA},\\
(1,1,-1,0)^{T}\,\,\,\mbox{for GC},\,\,\,\,\,\, & (0,-1,1,1)^{T}%
\,\,\,\mbox{for CG}
\end{array}
\right.
\end{equation}
where $+1,-1,0$ denote, respectively, an acceptor, a donor, and a missing
bond, that each of the four base pairs can form with an external ligand at
position $n$ on the DNA \cite{See76}. We also assume that the H-bonds formed
in the DNA-protein complex at the recognition site are known (this information
can be obtained from crystallographic analysis of the DNA-protein complex).
The protein can then be represented by a ($r\times4$) \emph{recognition
matrix} $R$ describing the pattern of H-bonds formed by the protein and the
DNA at the recognition site. The protein-DNA interaction energy is then
defined by counting the matching and unmatching bonds between the recognition
matrix and the DNA sequence at site $n$,
\begin{equation}
E(n)={\mathcal{\epsilon}}\;\;t\!r(R\cdot B_{n})\,,
\end{equation}
where $\mathcal{\epsilon}$ denotes each H-bond energy, $\,t\!r$ the trace, and
the dot refers to usual matrix multiplication. The DNA is thus viewed as a
one-dimensional vector lattice characterized by a rough on-site potential
$E(n)$, on which a random walker (a protein) moves, with rates (probability
per unit time)
\begin{equation}
\left\{
\begin{array}
[c]{lll}%
r_{n\rightarrow n^{\prime}} & = & \min{(1/2\,,\,1/2\,\exp{(-\beta\,\Delta
E_{n\rightarrow n^{\prime}})})},\\
r_{n\rightarrow n} & = & 1-r_{n\rightarrow n+1}-r_{n\rightarrow n-1}\,;
\end{array}
\right.  \label{eq:rate}%
\end{equation}
where $n^{\prime}=n\pm1$ and $\beta=1/k_{B}T$. Time is measured in one-step
time units ($t.u.$). An estimation for the lower bound of the time unit can be
obtained through simple hydrodynamic considerations \cite{Sch79,Bar04},
yielding $1\,t.u.\approx10^{-7}\;s$. The typical H-bond energy is of order of
a few kcal/mol, but in fact the actual ${\mathcal{\epsilon}}$ could be much
less due to screening introduced by the water layer around DNA.

The presence of an activation barrier for the translocation on neighboring
sites can be accounted for by introducing a uniform threshold energy level
$E_{t}$, so that
\begin{align}
\label{eq:threshold}\Delta E_{n \to n^{\prime}} =  &  \max[E_{t} - E(n),\,
E(n^{\prime})-E(n),\, 0]\,.
\end{align}
Note that the effective translocation barrier also depends on the position,
through the on-site energy. As a specific example, we consider the case of the
T7 RNA-polymerase sliding on the bacteriophage T7 DNA. For this case it is
known that the recognition site is the five bps sequence $GAGTC$ extending
from position -11 to -7 in the $T7$ promoter. The interaction matrix R can be
inferred from the crystallographic studies of Cheetam et al. \cite{Che99}, as
\begin{equation}
R = \left(
\begin{array}
[c]{rrrr}%
1 & 1 & 0 & 0\\
1 & -1 & 0 & 0\\
1 & 1 & 0 & 0\\
0 & 1/2 & 0 & 0\\
0 & 0 & 1/2 & 1
\end{array}
\right),
\end{equation}
where the presence of $1/2$ is due to one shared DNA-protein
 H-bond mediated by a
water molecule and therefore considered as two half bonds. 
\begin{table}[ptb]
\caption{ The short time sub-diffusive parameters $A$ and $b$ fitted in the
initial time interval $[0,100]$, and those characterizing the asymptotic
regime, $D_{\infty}$ and $b_{\infty}$, fitted in $t \in[8\, 10^{6}, \;
10^{7}]$. The equilibrium diffusion constant $D^{*}$ is estimated from the
\textit{mfpt} analysis. All values are obtained for $\beta{\mathcal{\epsilon}}
= 1$.}%
\label{table1}
\begin{tabular}
[c]{c|cc|cc|c}%
$E_{t}$ & $2A$ & $b$ & $2D_{\infty}$ & $b_{\infty}$ & $2D^{*}$\\\hline
$E_{min}$ & $0.82 \pm  2\%$ & $0.49 \pm  1\%$ & $4.4 \, 10^{-3} \pm  1\%$ &
$0.94 \pm  1\%$ & $4.4 \, 10^{-3}$\\
0 & $0.48 \pm  2\%$ & $0.56 \pm  1\%$ & $4.3 \, 10^{-3} \pm  1 \%$ & $0.93
\pm  1\%$ & $4.3 \, 10^{-3}$\\
$E_{max}$ & $0.04 \pm  3\%$ & $0.61 \pm  1\%$ & $0.25 \, 10^{-3} \pm  2\%$ &
$0.83 \pm  1\%$ & $0.2 \, 10^{-3}$\\
&  &  &  &  &
\end{tabular}
\end{table}


\section{The properties: subdiffusivity and crossover to normal diffusion}

Theoretically, one can easily calculate the stationary distribution of a
population of proteins on the energy landscape as $\rho_{\infty}(n)\propto
\exp{(-\beta E(n)})$, only dependent on the site energy and on temperature.
This implies that the energy minima, that correspond to the recognition sites,
will be in average the most populated. We then calculate the mean square
deviation from the average of the spatial displacement, $\langle\Delta
n^{2}\rangle=\sum_{i=1}^{N}(n_{i}(t)-n_{i}(0))^{2}$, where average over
initial positions and different histories (Monte-Carlo runs) is made. The
three cases $E_{t}=min[E(n)]\equiv E_{min}$, $E_{t}=0$ and $E_{t}%
=max[E(n)]\equiv E_{max}$ have been examined. In the limit $\beta
{\mathcal{\epsilon}}=0$ the linear diffusion is recovered, as one expects (the
limiting value $2D=1$ is obtained in the case $E_{t}=E_{min}$, i.e., for a flat
potential without thresholds). Nevertheless, in the finite temperature case,
we obtain large initial deviations from the normal diffusion behaviour. More
precisely, for all thresholds we find that at the initial stage the diffusion
displays anomalous \emph{ sub-diffusive} features, with
\begin{equation}
\langle n^{2}\rangle=2At^{b},\;\;\;b<1\label{eq9}%
\end{equation}
where $A$ and $b$ depend on the fixed threshold level. The appearance of the
initial subdiffusive regime is not surprising, and has been observed both for
random trap and random barrier potentials, see, e.g., \cite{Kehr_review}.
\ Our case in Eq.(\ref{eq:rate}), however, represents a mixture of these two
cases, for which to our knowledge, there are no studies for the initial time
behaviour. On the other hand, note that in Eq.(\ref{eq:rate}) the hopping
rates $r_{n\rightarrow n+1},r_{n\rightarrow n-1}$ are not random variables but
depend on the gradient of the energy landscape, $log(r_{n\rightarrow
n+1}/r_{n+1\rightarrow n})=(E_{n+1}-E_{n})/(k_{B}T)$. This has the important
consequence that in the continuous (Langevin) approximation of the process
(see, e.g., \cite{Bouchaud_review}), the effective potential $U$ stays
gaussian localized with the typical difference $U(n)-U(n-1)\approx\sqrt
{2}\sigma_{E}$ independent of $n$, $\sigma_{E}$ being the energy variance.
This is different from Sinai model where typical $U(n)$ grows with $n$ as
$\sqrt{n}$, this leading to anomalous $\langle x^{2}\rangle\sim(ln(t))^{4}$
behaviour . Since Sinai model is not applicable to our case, we will be using
in the following a rather crude approximation (\ref{eq9}) to describe the
crossover from initial subdiffusion to linear diffusion regime.A quantitative
characterization of the initial transient regime is given in Table
\ref{table1}, for the three values of $E_{t}$. The diffusion constant
$D_{\infty}$ for the three threshold levels is estimated from the linear fit
$\langle\Delta n^{2}\rangle=2Dt$ at large times $t\in\lbrack8\times
10^{6},\;10^{7}]$. We checked that an effective linear behaviour is roughly
reached by evaluating the parameter $b_{\infty}$ in the same range (see Table
\ref{table1}). Asymptotically, a standard diffusion is recovered (on the large
scale the potential roughness averages to zero). The asymptotic diffusion
constant $D$ decreases for increasing $\beta{\mathcal{\epsilon}}$. The initial
deviation from a random walk ($1-b$) and the time needed to reach the
asymptotic limit both increase with $\beta{\mathcal{\epsilon}}$; the typical
one-step time (or time unit $t.u.$) should be roughly, for real proteins, of
order of micro-second \cite{Bru02,Sch79}, thus giving crossover times up to
seconds corresponding to mean displacements up to hundreds bps (data not
shown; more details will be given elsewhere). A theoretical estimate of the
large time effective diffusion constant can be obtained from mean first
passage time (mfpt) analysis. According to Ref.~\cite{kampen}, for a discrete
one step process, such as the one considered here, the mfpt $T_{n_{0}}^{n}$ to go from
a referring position $n_{0}$ to position $n>n_{0}$ can be evaluated, once a
reflecting barrier is fixed in a position $a<n_{0}$, as
\begin{equation}
T_{n_{0}}^{n}=\sum_{i=n0}^{n}\frac{1}{r_{i\rightarrow i-1}\,\,\rho_{\infty
}(i)}\,\,\sum_{j=a}^{i}\rho_{\infty}(j).\label{eq13}%
\end{equation}
Note that $T_{n_{0}}^{n}$ depends on the threshold level $E_{t}$ through the
rate $r_{n\rightarrow n-1}$, according to Eq.~(\ref{eq:rate}). For
 large enough $T_{n_{0}}^{n}$,
\begin{equation}
\langle\Delta n^{2}\rangle\approx2DT_{n_{0}}^{n}
\label{eq14}%
\end{equation}
\begin{figure}[ptb]
\includegraphics[height=.42\textwidth,angle=270]{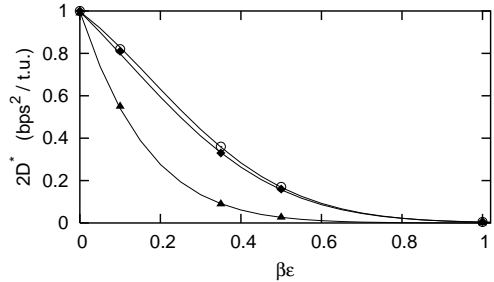}\caption{$2D^{\ast
}=\langle\Delta n^{2}\rangle/T_{n0}^{n}$ as a function of the adimensional
parameter $\beta{\mathcal{\epsilon}}$ (full lines), and the corresponding $2D$
directly evaluated by fitting the large time diffusion (symbols), for
corresponding different values of the threshold energy: $E_{t}=E_{min}$ (open
circles), $E_{t}=0$ (triangles) and $E_{t}=E_{max}$ (diamonds). Time is
measured in time units (t.u.), see text for details.}%
\label{fig1}%
\end{figure}Making the choice $a=0$, the theoretical diffusion constant
$D^{\ast}$ as a function of $\beta\mathcal{\epsilon}$ can be evaluated using
Eq.~(\ref{eq14}). The result is shown on Fig.~\ref{fig1} together with the
corresponding numerically evaluated diffusion constants. We observe an
excellent agreement. Note that the diffusion constant decreases exponentially
for $\beta{\mathcal{\epsilon}\rightarrow\infty}$ ( in practice, it is already
$\ll1$ for $\beta{\mathcal{\epsilon}}\approx1$) and the corresponding mfpt
exponentially increases in the same limit. This behaviour reflects the
divergence of the typical extent of the sub-diffusive transient, which becomes
more and more important as $\beta{\mathcal{\epsilon}}$ approaches $1$.

The model allows also to consider the possibility that very unfavorable
positions (with a large number of mismatches) could induce protein
conformational changes to an extent of not allowing the formation of any
H-bond, thus inducing a regime of \textquotedblleft free
sliding\textquotedblright\ \cite{von96}. A threshold energy level should in
this case separate \textit{reading} regions from \textit{free sliding}
regions, (where the DNA is seen as homogeneous chain). The energy landscape
should then be redefined to be homogeneous above this threshold: we will put
$E(n)=E_{sl}$ if $E(n)\geq E_{t}$, and refer to this second possibility as
\emph{\textquotedblleft two-state model\textquotedblright}. In this case, the
redefinition of the energy landscape leads to a faster diffusion, even if
still sub-diffusive, at small times. This effect is more evident for low
threshold values, i.e., as the energy redefinition involves an increasing
number of sites. Indeed, if a particle (protein) is located on a flat part of
the potential, it will start to diffuse freely with maximally possible
diffusion constant. Such particles contribute to fast diffusion at initial
time. After having slid freely for a certain time, however, a particle will
fall in $E<E_{t}$ region, and will be partially trapped in a potential well.
After a transient time, a subdiffusive behaviour similar to the previous case
is indeed reached, that converges, on larger times, to linear diffusion. A
detailed analysis of the \textquotedblleft two-state model\textquotedblright%
\ will be presented elsewhere. \begin{figure}[ptb]
\includegraphics[height=.42\textwidth,angle=270]{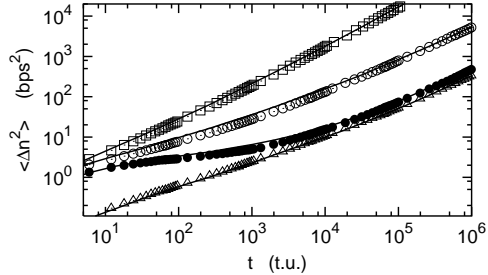}\caption{Dynamics
obtained on an artificial Gaussian energy landscape with $E_{min}=-N\epsilon$,
$E_{max}\approx N\epsilon/2$ (solid lines) compared to that obtained for the
T7 RNA-polymerase -- DNA interaction (symbols), for energy parameters:
$E_{t}=E_{min}$, $\beta\epsilon=0.5$ (squares); $E_{t}=E_{min}$,
$\beta\epsilon=1$ (open circles); $E_{t}=E_{max}$, $\beta\epsilon=1$
(triangles); \emph{\textquotedblleft two-state model\textquotedblright} with
$E_{t}=0$, $\beta\epsilon=1$ and $E_{sl}=E_{max}$ (full circles). Time is
measured in time units (t.u.), see text for details.}%
\label{fig2}%
\end{figure}

Thus, one sees a substantial deviation from random walk  during sliding phase
of a target search. In the next section, we address the question about the
generality of the presented results, in application to larger and more complex
proteins such as e.g. E. Coli RNA-polymerase, lac repressor, EcoRI and EcoRV,
i.e., for other H-bond reading enzymes.

\section{Generalization to other enzymes and binding mechanisms}

First of all, note that the dynamics of the proposed model  depends only on
the obtained energy profile, and that the most important parameter is the
single energy contribution $\epsilon$, that fixes the energy scale. This
quantity, though experimentally difficult to access, should nevertheless
depend only on the nature of the H-bond: one can thus reasonably expect it to
be roughly the same for all proteins. The actual threshold mechanism is also
unknown, but again we could reasonably expect that it depends on general
properties of the protein-DNA interaction, and does not vary in nature from
one protein to another.

What should represent the main difference between different proteins is
therefore the length of the recognition sequence \cite{noteHwa}, or, more
precisely, the number of bonds involved in the reading. This parameter  should
be adapted in order to mimic the sliding of different enzymes.

An examination of the whole set of possible hydrogen bonds that DNA bps can
form with external ligands \cite{See76,Nad99} shows that, among the $12$
possible H-bond sites exposed on the $4$ different bps, those that are in
central binding sites of the bases ($b_{n}[2]$ and $b_{n}[3]$) can induce both
matches or mismatches, while the external ones ($b_{n}[1]$ and $b_{n} [4]$)
are either matches or give zero contribution to the interaction energy. It is
thus possible to calculate explicitly the energy level distribution for a
generic enzyme looking for a total of $N$ matches with $N^{\prime}$ of them in
the two central binding sites of the bases. The only assumption made is that
the matches are uncorrelated, which turns out to be a reasonable approximation
for quasi-random DNA sequences. The resulting energy level distribution is a
convolution of two binomials that rapidly converges to a Gaussian as $N$ and
$N^{\prime}$ increase \cite{binom}. It is then easy to calculate the average
and standard deviation of the energy that result to be $\langle E\rangle
=(N-N^{\prime})\epsilon/2$ and $\sigma_{E}=(N+3N^{\prime})\epsilon/4$
respectively. The minimum and maximum energies of the resulting distributions
are given by $E_{min}=-N\epsilon$, $E_{max}=N^{\prime}\epsilon$.

This leads us to conclude that, for not too small values of $N$ (and
$N^{\prime} $), the energy level distribution is approximatively a Gaussian,
and its width just depends on $N$ and $N^{\prime}$ (or, alternatively,
$E_{min}$ and $E_{max}$ ). Note, furthermore, that if bonds on different
positions are equiprobable, $N^{\prime}$ should be roughly equal to $N/2$, so
that one ends with only one parameter. We can expect therefore that the energy
landscape for a generic sliding protein, and therefore the sliding motion
depends crucially on the number of H-bonds made at the recognition site.

We have tested the previous arguments by building an artificial energy
profile, with random levels distributed as to reproduce the original
distribution width and thus the original Gaussian shape. \begin{figure}[ptb]
\includegraphics[height=.42\textwidth,angle=270]{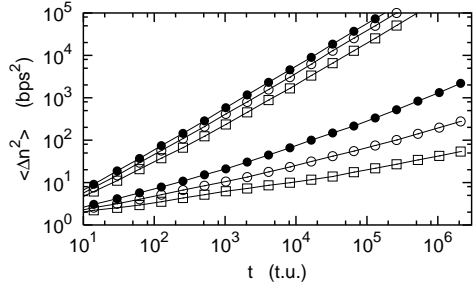}\caption{Dynamic
behaviour, obtained on the artificial Gaussian energy landscape, for $N=10$
(full circles), $N=14$ (open circles), $N=20$ (squares), with $\beta
{\mathcal{\epsilon}}=1$ (upper curves) or $0.2$ (lower curves). Time is
measured in time units (t.u.), see text for details.}%
\label{fig3}%
\end{figure}In Fig.~\ref{fig2}, simulations of the protein sliding motion on
the basis of this artificial energy landscape are compared with previous
results for different choices of the model parameters. Despite the certain
arbitrariness in the definition of artificial energy landscape, we obtain
essentially the same diffusive behaviour as for the true DNA case. In
Fig.~\ref{fig3} we depict the diffusive behaviour for three different values of
$N$, with $N^{\prime}=N/2$: as easily predicted, the asymptotic normal
diffusion slows down when the number of bonds increases. This parameter thus
affects the asymptotic diffusion regime as well as the initial subdiffusion
and the transition time.

\section{Conclusions}

In this paper we have considered the sliding motion of a protein on DNA by
means of a probabilistic model which includes the information about the base
sequence through the base pair reading interaction. In the case of the T7
RNA-polymerase we found that the protein executes a random motion which
deviates from the standard random walk dynamics usually assumed. We argued
that the same qualitative behaviour should be valid also for other types of
enzymes. The presence of an anomalous diffusion regime at the
early stages of the process speeds up the mobility of the protein facilitating
the target search. The overall diffusive behaviour of the sliding protein can
be characterized in terms of few parameters: the typical interaction energy
$\mathcal{\epsilon}$ associated with each DNA-protein bond, and the number $N$
of such bonds formed at the recognition site. We conclude that only few
parameters determine the overall diffusive behaviour of a sliding protein on
DNA: the typical interaction energy $\mathcal{\epsilon}$ associated with each
DNA-protein bond, and the number $N$ of such bonds formed at the recognition
site. One can therefore expect the same qualitative behaviour described here on
the example of $T7$ RNA-polymerase to be valid also for other types of enzymes
(if other kind of specific chemical bonds intervene in the recognition
mechanism, as e.g. water-bridges, minor groove H-bonds or hydrophobic contacts
\cite{Nad99,Kal02}, the corresponding energies should be evaluated and
included in the model; nevertheless, the number of specific bonds is strictly
a characteristic of each different enzyme-DNA interaction, and the diffusing
behaviour must still depend on this number).

We finally remark that the presence of additional sequence-dependent
interaction in the recognition process, such as the one involving geometrical
and elastic characteristics of the DNA, can also be included in our model.
This additional interaction, being sequence specific, would lead to a
redefinition of the energy landscape without effecting much the qualitative
results of the paper (they however are much more difficult to model due to the
scarcity of experimental data). In particular, discussed above anomalous
diffusion regime is robust with respect to changes of the energy landscape.
Therefore, the influence of the DNA sequence on the sliding motion of a
protein on DNA makes the standard random walk assumption for sliding phase of
the target search incorrect for a large set of parameters. Accounting for this
anomalous diffusive motion should be included in realistic description of the
sliding component of the target search in order to discriminate the relative
role of 1D sliding and 3D diffusion in the search process.

\acknowledgments MS and VP wish to acknowledge the Laboratoire de Physique
Theorique des Liquides, Universit\'{e} Paris VI, for hospitality and partial
support. MS and MB also acknowledge hospitality and partial support from the
Forschungszentrum Juelich, Germany.


\begin{thebibliography}{23}
\bibitem[{not()}]{noteHwa}
\bibinfo{note}{An interesting discussion on the relevance of the number of
  base-pairs involved in the recognition mechanism is also presented in
  Ref.~\cite{Ger02}.}
\bibitem[{bin()}]{binom}
\bibinfo{note}{Detailed calculations will be presented in a future work.}
\end{thebibliography}

\end{document}